\documentclass[
reprint,
superscriptaddress,
showpacs,
nofootinbib,
nobibnotes,
 amsmath,amssymb,
aps,
prb,
]{revtex4-2}
\usepackage{graphicx}
\usepackage{dcolumn}
\usepackage{appendix}
\usepackage{multirow}
\usepackage{placeins}
\usepackage[colorlinks=true,linkcolor=blue,anchorcolor=red,citecolor=blue,urlcolor=blue]{hyperref}

\begin{document}


\title{Measurement-induced phase transition in space }

\author{Jia-Qiang Li}
\affiliation{Guangdong Provincial Key Laboratory of Magnetoelectric Physics and Devices, Sun Yat-sen University, Guangzhou 510275, China}
\affiliation{School of Physics, Sun Yat-sen University, Guangzhou 510275, China}%
\author{Liang-Jun Zhai}%

\affiliation{School of Mathematics and Physics, Jiangsu University of Technology, Changzhou 213001, China}%

\author{Shuo Liu}
\email{sl6097@princeton.edu}
\affiliation{Department of Physics, Princeton University, Princeton, NJ 08544, USA}

\author{Shuai Yin}
\email{yinsh6@mail.sysu.edu.cn}
\affiliation{Guangdong Provincial Key Laboratory of Magnetoelectric Physics and Devices, Sun Yat-sen University, Guangzhou 510275, China}%
\affiliation{School of Physics, Sun Yat-sen University, Guangzhou 510275, China}%

\date{\today}

\begin{abstract}
Measurement-induced phase transitions (MIPTs) in monitored quantum circuits are usually characterized by preparing steady states at different uniform measurement probabilities. Here we introduce a spatial realization of the MIPT by imposing a deterministic measurement gradient in a single monitored Clifford chain. The resulting steady state contains coexisting volume-law, critical, and area-law regions, with the point $p(x)=p_c$ acting as a spatial critical cut. By scanning entanglement observables across this profile, we show that the transition is organized by a spatial scaling form. Although this structure is analogous to finite-time scaling in temporally driven MIPT, the spatial protocol has no Kibble-Zurek dynamics. Instead, the physical bounds $0\le p\le 1$ impose a finite linear window, producing cutoff-controlled asymptotic regimes whose fitted exponents provide direct access to the correlation-length exponent $\nu$. Our results establish spatially inhomogeneous measurements as a controlled route to engineer and probe measurement-induced criticality within a single steady state.
\end{abstract}

\maketitle

\section{Introduction}

Rapid progress in near-term quantum computers~\cite{arute_quantum_2019,cerezo_variational_2021,kim_evidence_2023,acharya_suppressing_2023,king_quantum_2023,bluvstein_logical_2024,miessen_benchmarking_2024,chiu_continuous_2025,acharya_quantum_2025,gao_establishing_2025,jiang_advancements_2025,berezutskii_tensor_2025,abanin_observation_2025,loschnauer_scalable_2025,he_experimental_2025} and quantum simulators~\cite{barreiro_open-system_2011,zhang_observation_2017,childs_toward_2018,tacchino_quantum_2020,scholl_quantum_2021,ebadi_quantum_2021,daley_practical_2022,fauseweh_quantum_2024} has made it possible to probe exotic phases of matter~\cite{keesling_quantum_2019,liu_discrete_2023,iqbal_topological_2024,guo_new_2024,wang_anderson_2025,liu_supersymmetry_2025,zeng_quantum_2019,lu_mixed-state_2023,feng_absence_2023}, nontrivial quantum dynamics~\cite{cao_entanglement_2019,choi_quantum_2020,alberton_entanglement_2021,feng_measurement-induced_2023,liu_universal_2023,kumar_boundary_2024,liu_noise-induced_2024,soto_garcia_resolving_2024,mochizuki_measurement-induced_2025,liu_symmetry_2025,liu_noisy_2025,xiao2026nonstabilizernessmpembaeffects}, and associated many-body protocols~\cite{fan_self-organized_2021,dupont_quantum_2022,Li_cross_2023,garratt_probing_2024,liu_entanglement_2024,Tikhanovskaya_universal_2024,guo_locally_2025,ha_absorbing_2025,qian_protect_2025}. These developments call for simple theoretical frameworks that isolate universal features of entanglement generation, information spreading, and measurement backaction in nonequilibrium many-body systems. Random quantum circuits provide such a framework, encompassing purely unitary circuits~\cite{nahum_quantum_2017,nahum_entanglement_2020,chen_emergent_2020,piroli_quantum_2021,sierant_measurement-induced_2022,foligno_temporal_2023,han_entanglement_2023,liu_symmetry_2024,sommers_zero-temperature_2024,chen_subsystem_2025,zhang_entanglement_2025,ares_entanglement_2025,turkeshi_quantum_2025,li_quantum_2025}, monitored circuits combining unitary gates with local measurements~\cite{li_quantum_2018,li_measurement-driven_2019,zhang_noise-induced_2022,liu_measurement-induced_2022,shkolnik_measurement_2023,akhtar_measurement-induced_2024,wang_driven_2024,wang_relaxation_2025,li_exact_2025,kelly_entanglement_2025,xu_multipartite_2025}, and non-unitary measurement-only circuits~\cite{lang_entanglement_2020,ippoliti_entanglement_2021,van_regemortel_entanglement_2021,sang_entanglement_2021,lavasani_topological_2021,sang_measurement-protected_2021,lavasani_measurement-induced_2021,klocke_topological_2022,lavasani_monitored_2023,klocke_majorana_2023,sriram_topology_2023,qian_steering-induced_2024,yu_measurement-driven_2025,klocke_entanglement_2025,yu_gapless_2025}.

A particularly striking phenomenon in monitored circuits is the measurement-induced phase transition (MIPT), which occurs at the level of individual quantum trajectories~\cite{vasseur_entanglement_2019,skinner_measurement-induced_2019,li_conformal_2021} and is therefore distinct from conventional phase transitions characterized by ensemble-averaged observables or ground-state properties~\cite{jian_measurement-induced_2020,bayat_entanglement_2022,fisher_random_2023}. For spatially uniform measurement probability $p$, weak measurements allow unitary dynamics to generate a volume-law entangled phase, whereas strong measurements suppress entanglement and produce an area-law phase. These two phases are separated by a critical point $p_c$ with logarithmic entanglement scaling~\cite{li_conformal_2021}. Motivated by its connections to quantum information dynamics, error-correction structures, and noisy intermediate-scale quantum devices, MIPT has attracted extensive theoretical~\cite{zhou_emergent_2019,gullans_scalable_2020,gullans_dynamical_2020,bao_theory_2020,buchhold_effective_2021,doggen_generalized_2022,morral-yepes_detecting_2023,sierant_entanglement_2023,poboiko_theory_2023,fava_monitored_2024,poboiko_measurement-induced_2024,poboiko_measurement-induced_2025,passarelli_nonstabilizerness_2025,doggen_evolution_2023,tan_exploring_2026,kelson-packer_phase_2025} and experimental attention~\cite{noel_measurement-induced_2022,ippoliti_fractal_2022,koh_measurement-induced_2023,hoke_measurement-induced_2023,feng_postselection-free_2025,Kamakari_experimental_2025}.

However, most studies of MIPT have focused on homogeneous monitored dynamics, where each steady state is prepared at a fixed measurement probability $p$ and the transition is reconstructed by comparing steady states at different values of $p$. Spatially inhomogeneous measurements provide a natural extension of this framework. Previous work has shown that quenched spatial disorder in the measurement pattern can modify the conventional MIPT and even drive the transition toward infinite-randomness criticality~\cite{zabalo_infinite-randomness_2023,shkolnik_measurement_2023}. A different route to probing MIPT criticality is provided by temporally driven monitored circuits, in which the measurement probability is swept through $p_c$ and the resulting dynamics obey a finite-time scaling form beyond the Kibble-Zurek mechanism~\cite{wang_driven_2024}. These developments raise a direct question: what remains of driven MIPT scaling if the sweep is not performed in time, but is instead encoded statically in space?

This question is also motivated by broader lessons from gradient-controlled quantum dynamics. In unitary systems, deterministic spatial gradients such as tilted potentials can reorganize transport and entanglement propagation, leading to phenomena including Stark many-body localization~\cite{schulz_stark_2019,morong_observation_2021,van_regemortel_entanglement_2021} and quantum many-body scars~\cite{serbyn_quantum_2021,sala_ergodicity_2020,su_observation_2023}. Local measurements provide a non-unitary mechanism for restricting entanglement growth through the quantum Zeno effect~\cite{misra_zenos_1977,li_quantum_2018}. It is therefore natural to ask how a deterministic spatial gradient in the measurement probability reshapes the entanglement structure of a monitored many-body state, and whether such a gradient can realize critical scaling directly in space.

In this work, we answer this question by introducing a spatial realization of the MIPT in monitored Clifford qubit chains (see Fig.~\ref{circuit}). We impose a linearly varying measurement probability $p(x)$, truncated by the physical bounds $0\le p(x)\le 1$, so that the system contains volume-law, critical, and area-law regions within a single steady state. The local value of $p(x)$ selects the entanglement regime at position $x$, while the point satisfying $p(x)=p_c$ acts as a spatial critical cut. We refer to the transition organized around this cut as the \emph{spatial MIPT}.

We show that the spatial MIPT is governed by a scaling form structurally analogous to finite-time scaling in temporally driven MIPT~\cite{wang_driven_2024}. The analogy, however, is only structural. In the spatial protocol there is no Kibble-Zurek time evolution; instead, the gradient has scaling dimension $r_s=1+1/\nu$, and the direction in which the subsystem is extended determines whether the critical cut probes the volume-law or area-law side. Moreover, because $p(x)$ is a probability, the bounds $0\le p(x)\le 1$ impose a finite linear window $\mathcal{A}_{\max}\sim R^{-1}$. This geometric cutoff controls the large-scale asymptotics and exposes the correlation-length exponent $\nu$ on both the volume-law side and the area-law side.

We verify this scaling picture using the subsystem entanglement entropy $S_{A}$, the ancilla entropy $S_Q$, the tripartite mutual information $I_3$, and the full spatial entanglement profile. We further confirm that the scaling is not tied to a special choice of the critical position or base cut by studying an alternative boundary-anchored protocol. Together, these results establish spatially inhomogeneous measurements as a controlled and resource-efficient route to probe measurement-induced criticality within a single monitored steady state.

The paper is organized as follows. In Sec.~\ref{scalingform}, we introduce the scaling forms for the standard, temporally driven, and spatial MIPT. In Sec.~\ref{Setup}, we describe the monitored circuit, the spatial measurement profile, and the observables used throughout this work. In Sec.~\ref{MIPT}, we verify the spatial scaling form and analyze the cutoff-controlled asymptotics. In Sec.~\ref{AA}, we present the boundary-anchored protocol as a robustness check, and discuss the experimental relevance. We conclude in Sec.~\ref{conclusion}.

\section{Scaling forms \label{scalingform}}

We first recall the finite-size scaling form of the standard MIPT with a spatially uniform measurement probability. Near the critical point $p_c$, the subsystem entanglement entropy obeys
\begin{align}
S(p,|A|)=\alpha\ln |A|+H(g|A|^{1/\nu}),\label{equil}
\end{align}
where $g=p-p_c$, $H$ is the scaling function, and $|A|$ is the subsystem size. For the monitored Clifford circuits considered here, the critical parameters are $\alpha=1.57(1)$ and $\nu=1.260(15)$~\cite{li_measurement-driven_2019,sierant_measurement-induced_2022}. Equation~\ref{equil} expresses that the distance from criticality is encoded in the dimensionless scaling combination $g|A|^{1/\nu}$, equivalently in the comparison between $|A|$ and the correlation length $\xi\sim |g|^{-\nu}$, while the entropy at $g=0$ grows logarithmically as $\alpha\ln |A|$.

For comparison, we recall the finite-time-scaling form of the temporally driven MIPT~\cite{wang_driven_2024}, where the measurement probability is swept linearly in time, $g(t)=R_t(t-t_c)$. The corresponding scaling form is
\begin{align}
S(R_t,g,|A|)=\alpha\ln |A|+F_t(R_t|A|^{r_t},g|A|^{1/\nu}),\label{temporalFTS}
\end{align}
where $F_t$ is the finite-time-scaling function and $r_t=z+1/\nu$ is the scaling dimension of the temporal driving rate. The appearance of the dynamical exponent $z$ reflects the fact that the temporal sweep is limited by critical slowing down.

For the spatial protocol studied in this work, the measurement probability varies in space rather than in time, $g(x)=p(x)-p_c=R(x-x_c)$, where $R$ is the spatial gradient and $x_c$ denotes the position satisfying $p(x_c)=p_c$. Since $R=\partial_x g$, the scaling dimension of $R$ is obtained by combining the scaling dimension of $g$ with one spatial derivative:
\begin{align}
r_s=1+\frac{1}{\nu}.
\end{align}
This leads to the spatial scaling ansatz
\begin{align}
S_A(R,g,|A|)=\alpha\ln |A|+F_s(R|A|^{r_s},g|A|^{1/\nu}).\label{scaling}
\end{align}

Equation~\ref{scaling} is structurally analogous to the temporal finite-time-scaling form in Eq.~\ref{temporalFTS}, but its physical origin is different. The spatial protocol concerns a steady state with a static measurement gradient and therefore contains no Kibble-Zurek time evolution. For the present Clifford circuits, $z=1$, so $r_s$ is numerically equal to $r_t$; nevertheless, the two exponents originate from different scaling operations. In the temporal problem, the temporal rate is tied to critical relaxation, whereas in the spatial problem the spatial gradient is tied to the spatial variation of the tuning parameter sampled by the subsystem. As shown below, this distinction becomes essential in the large-argument regime of the scaling function, where the bounded probability range $0\le p(x)\le 1$ imposes a finite spatial window and changes its asymptotic behavior.

\section{Setup and observables\label{Setup}}

\subsection{Circuit model and spatial measurement profile}

\begin{figure}[htbp]
    \includegraphics[width=1.0\linewidth]{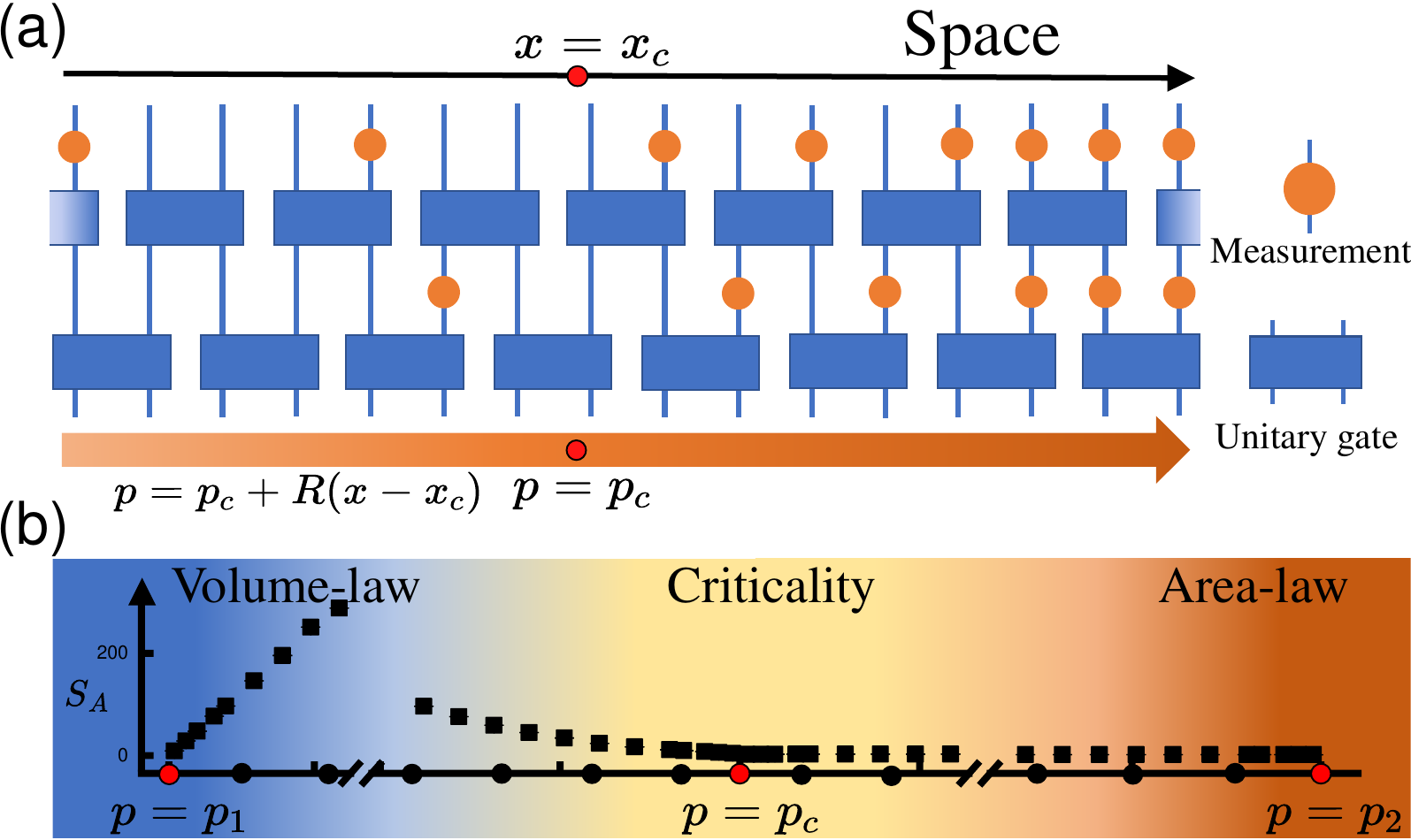}
    \caption{Schematic of the monitored Clifford circuit and the spatial MIPT. (a) The circuit consists of brickwork layers of two-site random Clifford gates, represented by blue rectangles; each gate layer is followed by a layer of single-site projective measurements, represented by orange circles. The measurement probability varies along the spatial direction according to the truncated linear profile $p(x)=\max\{0,\min\{1,p_c+R(x-x_c)\}\}$, illustrated by the gradient bar. (b) Entanglement structure generated by the spatial measurement profile. Within a single steady state, the chain contains volume-law, critical, and area-law regions. The point $p(x_c)=p_c$ acts as a spatial critical cut. The bipartite entanglement entropy $S_A$ is defined for a subsystem $A$ bounded by a base cut, marked by the red circle, and a second cut that is varied along the spatial direction.}
    \label{circuit}
\end{figure}

We consider a $(1+1)$D monitored random Clifford circuit, as illustrated in Fig.~\ref{circuit}. The system consists of a chain of $L$ qubits with periodic boundary conditions imposed on the unitary gates. Each time step contains two brickwork layers of two-site random Clifford gates,
\[
U(t)=\left[\prod_{x\,\text{odd}} U_{(x,x+1),\,2t}\right]
\left[\prod_{x\,\text{even}} U_{(x,x+1),\,2t+1}\right].
\]
After each gate layer, we apply local projective measurements in the $Z$ basis. At site $x$, the measurement projectors are $(\mathbf{I}\pm Z_x)/2$, and the measurement probability is $p(x)$. Measurement outcomes are sampled according to Born's rule, and the post-measurement state is normalized after each projection. In simulations, starting from the product state $|111\cdots\rangle$, we evolve the circuit for $2L$ time steps before measuring steady-state entanglement observables.

The key ingredient of the spatial MIPT is a deterministic spatial gradient in the measurement probability. Since $p(x)$ is a probability, the profile must be bounded within the physical interval $0\le p(x)\le 1$. We therefore use the truncated profile
\begin{align}
    p(x)=\max\{0,\min\{1,p_c+R(x-x_c)\}\}.
    \label{eq:probability_profile}
\end{align}
Here $x_c$ is the position at which $p(x_c)=p_c$. This construction produces a finite central window in which $p(x)=p_c+R(x-x_c)$, together with two saturated plateaus at $p=0$ and $p=1$. The scale of the central linear window is $\mathcal{A}_{\max}\sim R^{-1}$.

We retain periodic boundary conditions for the unitary layers in order to avoid hard geometric boundaries that would directly suppress entanglement growth. For the truncated measurement profile in Eq.~\eqref{eq:probability_profile}, periodicity produces a discontinuity in the measurement profile where the two ends of the chain are connected. This discontinuity has little influence on the critical scaling studied below because it is separated from the critical region by the saturated plateaus. 

\subsection{Subsystem partitions and observables}
\begin{figure}[htbp]
\includegraphics[width=1.0\linewidth]{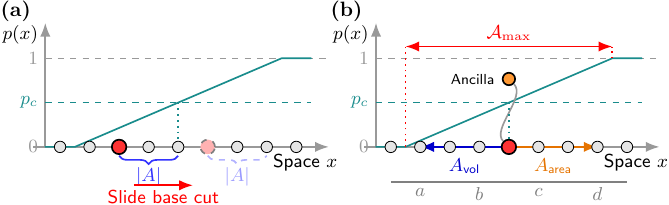}
\caption{Schematic illustrations of subsystem partitions. The central dashed line marks the critical position $x_c$, where $p(x_c)=p_c$, and the red circle denotes the base cut, which fixes one boundary of the subsystem. (a) For the entanglement distribution scaling, the subsystem size $|A|$ is fixed while the base cut slides along the spatial gradient, allowing the subsystem to sample different local values of $g=p-p_c$. (b) For the critical scaling, the base cut is fixed at $x_c$, and the second cut is moved either leftward into the volume-law side, defining $A_{\text{vol}}$, or rightward into the area-law side, defining $A_{\text{area}}$. The finite central linear window of the measurement profile defines the truncation scale $\mathcal{A}_{\max}$. The same base-cut convention is used for the ancilla entropy $S_Q$, while $I_3$ is evaluated using a four-segment partition of the chain.}
    \label{cut}
\end{figure}

The spatial scaling form in Eq.~\ref{scaling} depends not only on the subsystem size but also on how the subsystem is positioned relative to the spatial critical cut. We therefore use two complementary partition protocols, summarized in Fig.~\ref{cut}.

First, to probe the entanglement distribution across the gradient, we fix the subsystem size $|A|$ and slide the subsystem along the chain [Fig.~\ref{cut}(a)]. The base cut of each subsystem provides the local reference position. The measurement probability at this base cut defines the local tuning parameter $g=p-p_c$. At fixed $R|A|^{r_s}$, data for different subsystem sizes should then collapse as functions of the scaling variable $g|A|^{1/\nu}$.

Second, to probe the critical scaling, we anchor the base cut at the spatial critical point $p(x_c)=p_c$ and vary the subsystem size by moving the second cut [Fig.~\ref{cut}(b)]. Moving the second cut toward the $p<p_c$ side makes the subsystem extend into the volume-law region, while moving it toward the $p>p_c$ side makes the subsystem extend into the area-law region. This directional dependence is a distinctive feature of the spatial MIPT: the two sides of the same critical cut probe different entanglement regimes and therefore exhibit different large-scale asymptotic behaviors.

In addition to the subsystem entanglement entropy $S_A$, we use two standard diagnostics of measurement-induced criticality: the ancilla entropy $S_Q$ and the tripartite mutual information $I_3$~\cite{gullans_scalable_2020,zabalo_critical_2020}. To obtain $S_Q$, an ancilla qubit is coupled to the qubit adjacent to the base cut, and we monitor the entropy of the ancilla after the circuit reaches the steady state. We define
\begin{align}
    S_Q=-\operatorname{Tr}\rho_Q\ln\rho_Q,
    \label{eq:ancilla_entropy}
\end{align}
where $\rho_Q$ is the reduced density matrix of the ancilla qubit. This quantity diagnoses whether local quantum information remains encoded in the monitored many-body state.

The tripartite mutual information is obtained by dividing the chain into four equal segments labeled $a$, $b$, $c$, and $d$. It is defined as
\begin{align}
    I_3 \equiv S_{a} + S_{b} + S_{c}
    - S_{a \cup b} - S_{a \cup c} - S_{b \cup c}
    + S_{a \cup b \cup c},
    \label{eq:tripartite_mutual_information}
\end{align}
where $S_X$ denotes the von Neumann entropy of subsystem $X$. While $S_A$ probes bipartite entanglement across a chosen cut, $S_Q$ and $I_3$ provide complementary information about the encoding of quantum information and the multipartite entanglement structure near the spatial critical region.
\section{Spatial MIPT \label{MIPT}}
\subsection{Verification of the scaling form and cutoff-controlled asymptotics}
\begin{figure}[htbp]
\includegraphics[width=1.0\linewidth]{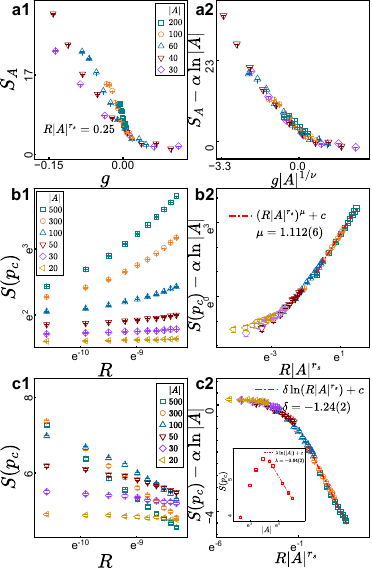}
    \caption{
    Verification of the spatial scaling form in Eq.~\eqref{scaling}. For fixed $R|A|^{r_s}=0.25$, panel (a1) shows the subsystem entropy $S_A$ as a function of $g$ for different subsystem sizes $|A|$, while panel (a2) presents the corresponding collapse after subtracting the critical logarithmic contribution. Panels (b) and (c) examine the critical scaling of $S_A(p_c)$. Panel (b1) plots $S_A(p_c)$ versus $R$ on the $p<p_c$ side for different $|A|$, and panel (b2) shows the associated collapse. The red dashed line in panel (b2) is a power-law fit with exponent $\mu=1.112(6)$. Panel (c1) shows $S_A(p_c)$ versus $R$ on the $p>p_c$ side, and panel (c2) presents the corresponding collapse. The semi-logarithmic fit gives slope $\delta=-1.24(1)$. The inset in panel (c2) shows $S_A(p_c)$ versus $|A|$ for $R=0.00025$, with slope $\lambda=-0.64(2)$.}
    \label{relation}
\end{figure}
We first verify the entanglement distribution scaling in the spatial MIPT. Because the transition unfolds across space, the position of the base cut naturally defines its distance from the spatial critical cut. When the subsystem size is sufficiently small compared with the total system length, we use its left boundary as the local reference point, as depicted in Fig.~\ref{cut}(a). This choice maps the spatial coordinate of the base cut to the local tuning parameter $g=p-p_c$, while the subsystem size $|A|$ sets the finite-size cutoff for the local critical behavior. We use a qubit chain of length $L=1024$, which is sufficiently large to suppress finite-size effects in the spatial scaling behavior while remaining computationally efficient for extensive numerical simulations.

For this test, we keep the scaling combination $R|A|^{r_s}$ fixed, which couples the subsystem size $|A|$ to the gradient $R$. Sliding the subsystem along the chain allows its base cut to sample a continuous sequence of local probabilities $p$ and thus the local tuning parameter $g=p-p_c$. Under this constraint, Eq.~\ref{scaling} reduces to 
\begin{align}
S_A=\alpha\ln |A|+F_0(g|A|^{1/\nu}).
\end{align} 
Thus, after subtracting the critical logarithmic contribution, $S_A-\alpha\ln |A|$ should collapse when plotted against $g|A|^{1/\nu}$. As shown in Fig.~\ref{relation}(a1), the unscaled entropies are well separated near $g=0$, while the rescaled data collapse onto a single curve in Fig.~\ref{relation}(a2), confirming the scaling form. Note that, to ensure a controlled scaling analysis, we restrict the data to the linear gradient region, explicitly excluding points that extend into the truncated plateaus ($p=0$ or $p=1$) to prevent boundary artifacts in the scaling collapse.

We next examine the critical scaling at the spatial critical point. At $g=0$, Eq.~\ref{scaling} becomes
\begin{align}
    S_A(p_c)=\alpha\ln |A|+F_1(R|A|^{r_s}).
\end{align}
Plotting $S_A(p_c)-\alpha\ln |A|$ as a function of $R|A|^{r_s}$ yields the collapses shown in Figs.~\ref{relation}(b2) and \ref{relation}(c2), corresponding to the volume-law and area-law sides, respectively. 
If the second cut moves toward the $p<p_c$ side, the subsystem increasingly samples the volume-law region; if it moves toward the $p>p_c$ side, it samples the area-law region. This is the spatial counterpart of the initial-state dependence in temporally driven MIPT~\cite{wang_driven_2024}, but here the distinction is selected by the direction of the subsystem cut rather than by the initial state.

The asymptotic behavior also differs from the temporally driven case. In Ref.~\cite{wang_driven_2024}, the temporal protocol gives a volume-law branch with $S(p_c)\propto R^{1/r_t}$ and an area-law branch with $S(p_c)\sim-(\alpha/r_t)\ln R$; the quantity $-\alpha/r_t$ is therefore a logarithmic slope, not a power-law exponent. These temporal forms follow from a Kibble-Zurek correlation length generated dynamically by the sweep rate. In the spatial protocol, however, the probability bounds $0\le p(x)\le 1$ impose a finite linear window
\begin{align}
    \mathcal{A}_{\max}\sim R^{-1}.
\end{align}
Once the subsystem probes this window, the large-$R|A|^{r_s}$ branch becomes cutoff-controlled rather than dynamically controlled: increasing $|A|$ further no longer samples a longer linear gradient, but instead reaches one of the truncated plateaus. Indeed, along $|A|\sim \mathcal{A}_{\max}$,
\begin{align}
    u\equiv R|A|^{r_s}\sim \mathcal{A}_{\max}^{-1}\mathcal{A}_{\max}^{1+1/\nu}
    =\mathcal{A}_{\max}^{1/\nu}.
\end{align}
On the volume-law side, the entropy is extensive in the available spatial window, $S_A\sim\mathcal{A}_{\max}$, up to subleading logarithmic and nonuniversal contributions. If the collapse function behaves as $F_1(u)\sim u^{\mu}$, then
\begin{align}
\left(\mathcal{A}_{\max}^{1/\nu}\right)^{\mu}\sim\mathcal{A}_{\max},
\end{align}
which gives
\begin{align}
    \mu=\nu.
\end{align}
This explains why the fitted exponent $\mu=1.112(6)$ in Fig.~\ref{relation}(b2) is much closer to $\nu\simeq1.26$ than to the temporal value $1/r_t\simeq0.558$. The remaining deviation from $\nu$ is expected, since the available scaling window is limited by both finite system size and the physical truncation of $p(x)$.

On the area-law side, the relevant length is the local correlation length associated with the tuning parameter at the outer cut, $g_A\sim R|A|$. Physically, extending the subsystem to the $p>p_c$ side moves the outer cut into a region where measurements suppress entanglement over a finite correlation length rather than allowing volume-law growth. Thus
\begin{align}
    \xi_A\sim g_A^{-\nu}\sim (R|A|)^{-\nu},
\end{align}
and, after absorbing nonuniversal additive constants, the entropy scales logarithmically as
\begin{align}
    S_A\sim \ln \xi_A\sim-\nu\ln R-\nu\ln |A|.
\end{align}
Writing the asymptotic correction as $F_1(u)\sim\delta\ln u$ gives
\begin{align}
    S_A=\delta\ln R+(\alpha+\delta r_s)\ln |A|.
\end{align}
 Matching the coefficient of $\ln R$ yields
\begin{align}
    \delta=-\nu,
\end{align}
consistent with the fitted slope $\delta=-1.24(1)$ in Fig.~\ref{relation}(c2), and the residual $|A|$ dependence has slope $\lambda=-0.64(2)$, close to
\begin{align}
    \lambda=\alpha-\nu-1\simeq -0.69.
\end{align}

The spatial asymptotics therefore do not indicate a different universality class; rather, they reflect how the finite spatial window reorganizes the scaling function and exposes the correlation-length exponent $\nu$ through both the power-law branch on the volume-law side and the logarithmic branch on the area-law side.

\subsection{Other observables and the entanglement distribution}
\begin{figure}[htbp]
\includegraphics[width=1.0\linewidth]{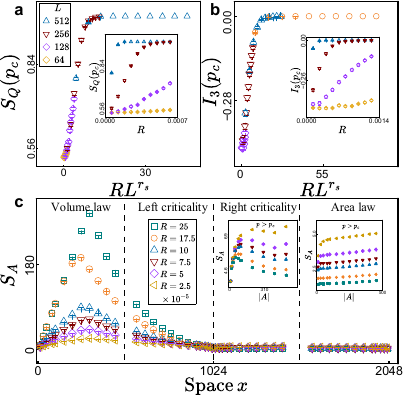}
    \caption{Critical scaling of the ancilla entropy $S_Q$ and tripartite mutual information $I_3$, together with the spatial entanglement profile. Panels (a) and (b) show the rescaled collapses of $S_Q$ and $I_3$, respectively; the insets show the corresponding unscaled critical data. Panel (c) shows the spatial coexistence of volume-law and area-law regions across the central critical point. The outer subpanels of (c) display the subsystem entanglement entropy far from $p_c$ for $p < p_c$ and $p > p_c$, respectively, revealing the volume-law and area-law entanglement structures characteristic of the spatial MIPT.}
    \label{SQ_I3}
\end{figure}
Using the setup in Fig.~\ref{cut}, we also examine the critical scaling of the ancilla entropy $S_Q$ defined in Eq.~\eqref{eq:ancilla_entropy} and the tripartite mutual information $I_3$ defined in Eq.~\eqref{eq:tripartite_mutual_information}. These observables probe complementary aspects of the transition: $S_Q$ diagnoses whether a local reference qubit remains entangled with the monitored system, while $I_3$ captures the multipartite structure of information shared among distant spatial regions. Since both are dimensionless critical diagnostics, their leading critical behavior is controlled by the same spatial scaling variables but does not contain the additive logarithmic term present in the subsystem entropy. The spatial scaling forms are
\begin{align}
    S_Q(R,L,g)&=G_Q(RL^{r_s},gL^{1/\nu}),\\
    I_3(R,L,g)&=K(RL^{r_s},gL^{1/\nu}),
\end{align}
so that at criticality $S_Q(p_c)=G_{Q1}(RL^{r_s})$ and $I_3(p_c)=K_1(RL^{r_s})$. The numerical results in Fig.~\ref{SQ_I3}(a,b) are consistent with these forms, showing that the spatial scaling framework is not restricted to the subsystem entropy $S_A$. For $I_3$, the approach to area-law behavior at large $R$ and $L$ is expected because, in the four-segment partition used for $I_3$, segment $c$ lies on the $p>p_c$ side, reflecting the spatial dependence of the entanglement regime.

To resolve the full spatial profile of the transition and observe the flat truncation plateaus, which are typically obscured in smaller systems, we use a large system size $L=2048$ with the critical point at the center, $x_c=1024$. By partitioning the system from the left ($x=0$) and right ($x=L$) boundaries, we probe the entanglement deep inside the volume-law and area-law regions, as shown in Figs.~\ref{circuit}(b) and \ref{SQ_I3}(c). This provides a direct spatial image of the coexistence implied by the scaling analysis. Figure~\ref{SQ_I3}(c) exhibits qualitatively distinct behavior on the two sides of the critical point. Far from the spatial critical point, the leftmost panel shows volume-law growth of the subsystem entropy with $|A|$ in the $p<p_c$ region. When $R$ is small,
the measurement probability at the left cut point is close to $p_c$. In this case, $S_A$ exhibits a logarithmic dependence on $|A|$, consistent with the critical entanglement scaling reported in Ref.~\cite{li_measurement-driven_2019}; for larger $R$, $S_A$ crosses over toward a linear dependence at small $|A|$. The two central panels probe the near-critical entanglement structure by fixing one cut at $p_c$ and varying the subsystem size, as shown in Fig.~\ref{cut}(b). On the $p>p_c$ side, as shown in the rightmost panel,
the subsystem entropy follows logarithmic growth for small $R$, whereas for large $R$ it becomes nearly independent of $|A|$ and approaches the area-law limit. Overall, these results provide a unified picture of the spatial MIPT, encompassing the universal scaling form, the coexistence of distinct entanglement regimes, and the spatial distribution of entanglement across the gradient.
\section{An alternative protocol for the spatial MIPT \label{AA}}
\subsection{Verification of the scaling form\label{AB}}
\begin{figure}[htbp]
    \includegraphics[width=1.0\linewidth]{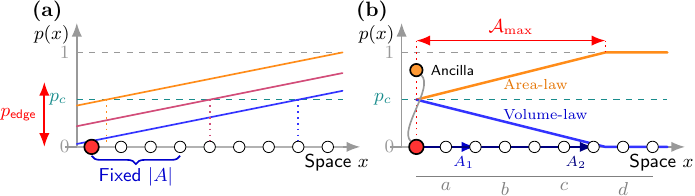}
    \caption{Sketch of the alternative boundary-anchored protocol. The base cut (red circle) is fixed at the left edge of the chain. Panel (a) illustrates the entanglement distribution scaling, where the subsystem size $|A|$ is fixed and the boundary value $p_{\text{edge}}$ is swept. This sweep translates the gradient profile vertically and shifts the critical point $x_c$ along the chain. Panel (b) depicts the critical scaling, where the boundary probability is fixed at $p_c$ and the subsystem size is varied by shifting the second cut into the bulk to form subsystems such as $A_1$ and $A_2$. The bounded profile sets the finite linear window $\mathcal{A}_{\max}$. The chain is partitioned into four equal segments labeled $a$, $b$, $c$, and $d$ to extract $I_3$, and an ancilla is coupled to the left edge to extract $S_Q$.}
    \label{cut1}
\end{figure}
\begin{figure}[htbp]
    \includegraphics[width=1.0\linewidth]{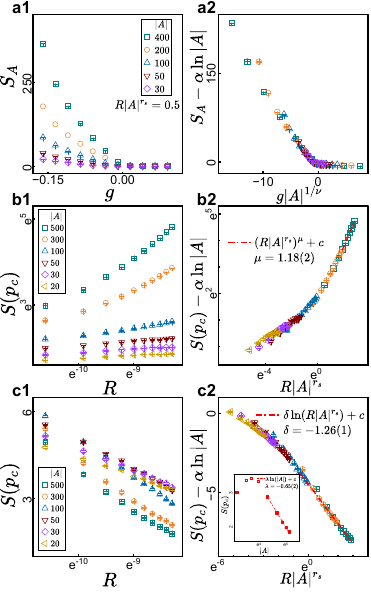}
    \caption{Verification of the scaling form in the boundary-anchored protocol. For fixed $R|A|^{r_s}=0.5$, panel (a1) shows the subsystem entropy $S_A$ as a function of $g$ for different subsystem sizes $|A|$, while panel (a2) presents the corresponding rescaled data. Panels (b) and (c) examine the critical scaling for the setup in Fig.~\ref{cut1}(b). Panel (b1) plots $S_A(p_c)$ versus $R$ for $p(x)=p_c-R(x-x_c)$, and panel (b2) displays the associated collapse. The red dashed line in panel (b2) is a power-law fit with exponent $\mu=1.18(2)$. Panel (c1) shows $S_A(p_c)$ versus $R$ for $p(x)=p_c+R(x-x_c)$, and panel (c2) presents the corresponding collapse. The semi-logarithmic fit gives slope $\delta=-1.26(1)$. The inset in panel (c2) shows $S_A(p_c)$ versus $|A|$ for $R=0.00025$, with slope $\lambda=-0.65(2)$.}
    \label{scale}
\end{figure}

We further examine whether the spatial location of $p_c$ affects the observed scaling. This is an important consistency check because, in the first protocol above, the critical point and the base cut are chosen at the middle of the chain. To show that the scaling does not rely on this special geometry, we design an alternative configuration in which the base cut is anchored at the left edge of the chain. For the entanglement distribution scaling, we fix the subsystem size adjacent to this boundary and sweep the boundary probability $p_{\text{edge}}$. At fixed gradient $R$, this sweep translates the entire profile and shifts the critical point $x_c$ across the system, as shown in Fig.~\ref{cut1}(a). The resulting sequence of local probabilities provides the scaling variable $g=p-p_c$. Figure~\ref{scale}(a) shows that the data collapse according to $S_A=\alpha\ln |A|+G_0(g|A|^{1/\nu})$, confirming that the entanglement distribution scaling is independent of the absolute position of the base cut.

We then test the critical scaling in the edge-anchored geometry by fixing the boundary probability at $p_c$ [Fig.~\ref{cut1}(b)]. The subsystem is extended either into the volume-law side, described by $p=p_c-R(x-x_c)$, or into the area-law side, described by $p=p_c+R(x-x_c)$. This protocol isolates the two directions away from the same critical boundary cut and therefore provides a clean check of the directional asymmetry discussed above. The data in Figs.~\ref{scale}(b,c) again collapse according to the critical form $S_A(p_c)=\alpha\ln |A|+G_1(R|A|^{r_s})$. On the volume-law branch, the fitted exponent $\mu=1.18(2)$ is consistent with the cutoff-controlled expectation $\mu\simeq\nu$. On the area-law branch, the logarithmic slope $\delta=-1.26(1)$ agrees with $\delta\simeq-\nu$, and the residual $|A|$ dependence has slope $\lambda=-0.65(2)$, which is consistent with the above results.
Thus, translating the critical point does not change the spatial scaling. The alternative protocol confirms that the cutoff-controlled asymptotics are an intrinsic consequence of the bounded measurement profile rather than a peculiarity of the centered geometry or the particular choice of base cut.

\subsection{Critical entanglement diagnostics}

\begin{figure}[htbp]
\includegraphics[width=1.0\linewidth]{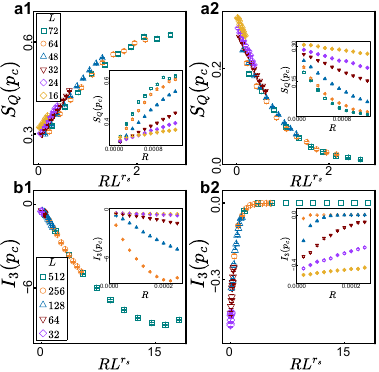}
    \caption{Critical ancilla entropy $S_Q$ and tripartite mutual information $I_3$ in the boundary-anchored protocol. Panels (a1) and (a2) show $S_Q$, while panels (b1) and (b2) show $I_3$. The left and right panels correspond to the volume-law and area-law sides, respectively. Insets show the unscaled critical data as functions of $R$ for different system sizes $L$, and the main panels show the corresponding collapses as functions of $RL^{r_s}$.}
    \label{SQ}
\end{figure}

Figures~\ref{SQ}(a) and \ref{SQ}(b) present the results of $S_Q(p_c)$ and $I_3(p_c)$, respectively. The left panels correspond to $p=p_c-R(x-x_c)$ and probe the volume-law side, while the right panels correspond to $p=p_c+R(x-x_c)$ and probe the area-law side. In Fig.~\ref{SQ}(a1), $S_Q(p_c)$ exhibits pronounced finite-size effects before rescaling, while the main panel shows a clear collapse with only small deviations for the smallest systems. In Fig.~\ref{SQ}(a2), $S_Q(p_c)$ decreases with increasing $R$, consistent with the decay of the ancilla entropy in the area-law regime~\cite{gullans_scalable_2020}; the rescaled curves collapse and approach zero at large $RL^{r_s}$. The tripartite mutual information shows analogous behavior. Figure~\ref{SQ}(b1) confirms the scaling collapse on the volume-law side, while Fig.~\ref{SQ}(b2) shows rapid decay toward zero on the area-law side at large $RL^{r_s}$. These results demonstrate that the same spatial scaling variables organize both bipartite and multipartite diagnostics, further supporting the spatial scaling framework beyond the subsystem entanglement entropy.

\subsection{Discussion}

Before concluding, we discuss the potential experimental relevance of the spatial protocol. On programmable platforms where measurement locations and probabilities can be controlled, the deterministic profile $p(x)$ can be implemented by applying the same monitored circuit with a site-dependent measurement probability. This requires no change to the underlying unitary circuit and avoids preparing a separate uniform-$p$ steady state for every value of the tuning parameter. Instead, a single steady state provides a spatial cross section of the MIPT, containing volume-law, near-critical, and area-law regions. This structure may reduce run-to-run calibration overhead in experiments where MIPTs have been observed~\cite{koh_measurement-induced_2023,hoke_measurement-induced_2023,noel_measurement-induced_2022}. The deeply measured area-law region can also serve as an \textit{in-situ} low-entanglement reference for monitoring background errors and hardware drift~\cite{majumder_real-time_2020,Singh_mid-circuit_2023,das_case_2019}.

The spatial protocol also suggests a practical route to reducing the resources needed to extract critical behavior. In a uniform protocol, the transition is reconstructed by comparing many steady states prepared at different values of $p$. In contrast, the spatial gradient encodes the tuning parameter directly in position, while the finite linear window $\mathcal{A}_{\max}\sim R^{-1}$
sets the largest length scale over which the local critical profile is sampled. As a result, the correlation-length exponent $\nu$ can be inferred from restricted spatial windows near the critical cut rather than from the full system. This cutoff-controlled structure does not remove the intrinsic sampling and post-selection challenges of monitored quantum dynamics, but it can mitigate them by focusing the required measurements on local subsystems whose effective size is set by $\mathcal{A}_{\max}$. Combined with scalable probes such as cross-entropy benchmarking~\cite{Li_cross_2023,Tikhanovskaya_universal_2024,Kamakari_experimental_2025}, this provides a promising strategy for studying measurement-induced criticality beyond Clifford circuits.

\section{Conclusion \label{conclusion}}
In this work, we proposed a protocol to realize the spatial MIPT and systematically verified its entanglement distribution and critical scaling forms through numerical simulations of random Clifford circuits. By imposing a deterministic spatial measurement gradient, we demonstrated that distinct entanglement regimes coexist within a single steady state, with the point satisfying $p(x)=p_c$ acting as a spatial critical cut between them. The critical scaling behavior of several observables, including the subsystem entanglement entropy, the ancilla entropy $S_Q$, and the tripartite mutual information $I_3$, is consistently captured by the spatial scaling framework based on the scaling combinations $R|A|^{r_s}$ and $RL^{r_s}$.

An important theoretical implication is that the bounded measurement probability imposes a finite spatial window, converting the large-scale asymptotics into a cutoff-controlled regime. This geometric constraint reorganizes the scaling function so that the fitted power-law exponent on the volume-law side and the logarithmic slope on the area-law side are governed by the correlation-length exponent $\nu$. In this sense, the spatial protocol separates the spatial scaling dimension from temporal Kibble-Zurek dynamics while preserving the underlying MIPT universality. By investigating an alternative boundary-anchored configuration, we further confirmed that the critical scaling is robust under translations of the spatial critical point $p_c$. Our results deepen the understanding of criticality in spatially inhomogeneous monitored circuits and establish a resource-efficient route for probing measurement-induced quantum phenomena.
\section{Acknowledgments}
J.-Q. L. and S. Y. are supported by the National Natural Science Foundation of China (Grant No. 12222515), the Research Center for Magnetoelectric Physics of Guangdong Province (Grant No. 2024B0303390001), the Guangdong Provincial Key Laboratory of Magnetoelectric Physics and Devices (Grant No. 2022B1212010008), and the Science and Technology Projects in Guangzhou City (Grant No. 2025A04J5408). S. L. was supported by the Gordon and Betty Moore Foundation through Grant No. GBMF8685 toward the Princeton theory program, the Gordon and Betty Moore Foundation’s EPiQS Initiative (Grant No. GBMF11070), the Global Collaborative Network Grant at Princeton University, the Simons Investigator Grant No. 404513, the Princeton Global Network, the NSF-MERSEC (Grant No. MERSEC DMR 2011750), the Simons Collaboration on New Frontiers in Superconductivity (Grant No. SFI-MPS-NFS-00006741-01 and No. SFI-MPS-NFS-00006741-06), the Princeton Catalysis Initiative, and the Schmidt Foundation at Princeton University. L.-J. Z. is supported by the National Natural Science Foundation of China (Grant No. 12274184).
\bibliography{Space}

\end{document}